\documentclass[10pt,twocolumn]{IEEEtran}
\usepackage[T1]{fontenc}
\usepackage{amsfonts}
\usepackage{amssymb,amsmath,color,graphicx}
\usepackage{subfigure}
\usepackage{cite}
\usepackage{pdfpages}
\usepackage{comment}
\usepackage{verbatim}
\usepackage{algorithm}
\usepackage{algorithmic}
\usepackage{placeins}
\usepackage{xfrac}
\usepackage{color,soul}
\usepackage{graphics} 
\usepackage{epsfig} 
\usepackage{times}
\usepackage{amsmath}
\usepackage{fix-cm}
\usepackage[T1]{fontenc}

\usepackage{epstopdf}

\usepackage{amssymb,amsmath,color,graphicx}
\usepackage{subfigure}
\usepackage{cite}
\usepackage{verbatim}
\usepackage{algorithm}
\usepackage{algorithmic}
\usepackage{setspace}
\usepackage{float}
\usepackage[T1]{fontenc}
\usepackage[english]{babel}
\usepackage{amssymb,amsmath,color,graphicx}
\usepackage{subfigure}
\usepackage{cite}
\usepackage{verbatim}
\usepackage{algorithm}
\usepackage{algorithmic}
\usepackage{setspace}
\usepackage{float}
\usepackage{dsfont}
\usepackage{bbold}
\usepackage{dsfont}
\usepackage{xr}
\usepackage{tabu}
\usepackage{tabularx}
\usepackage{longtable}
\usepackage{slashbox}
\usepackage[utf8]{inputenc}
\usepackage{pifont}
\usepackage{wasysym}
\usepackage{amssymb}
\usepackage{amsmath,bm}
\allowdisplaybreaks[1]
\usepackage{color}
\begin{document}

\title{\fontsize{18}{18} \selectfont \vspace{-4mm} A Decentralized  Framework for   Real-Time Energy  Trading in\\\vspace{1mm}  Distribution Networks with Load and Generation Uncertainty}
\author{\vspace{-2mm}\IEEEauthorblockN{Shahab Bahrami and M. Hadi Amini}

\IEEEauthorblockA{ email:  bahramis@ece.ubc.ca, amini@cmu.edu \vspace{-10.5mm}}
 }

\maketitle

\begin{abstract}

The proliferation of small-scale renewable generators and price-responsive loads makes it a challenge for distribution network operators (DNOs) to schedule the controllable loads of the load aggregators and the generation  of the generators in real-time. Additionally,  the high computational burden and violation of the entities' (i.e., load aggregators’ and generators’) privacy make a centralized framework impractical.  In this paper, we propose a \textit{decentralized} energy trading algorithm that can be executed by the entities in a real-time fashion.  To address the  privacy issues, the DNO  provides the entities with proper \textit{control signals} using the Lagrange relaxation technique to motivate them towards  an operating point with  maximum  profit  for entities. To deal with uncertainty issues, we propose a probabilistic load model and robust framework for renewable generation.  The performance of the proposed algorithm is evaluated on an  IEEE 123-node test feeder. When compared with a benchmark of not performing load management for the aggregators,  the proposed  algorithm   benefits both the load aggregators and generators by increasing their profit by 17.8$\%$ and  10.3$\%$, respectively.  When compared with a centralized approach, our algorithm converges to the solution of the DNO's centralized problem with a significantly lower running time  in $50$ iterations per time slot.

\vspace{1mm}
\noindent \textit{Keywords}: price-responsive load, generation uncertainty, distributed algorithm, trading market.
\end{abstract}

\vspace{-5mm}
\section{Introduction} \label{sec:introduction}
\vspace{-1mm}
One goal of the emerging smart grid is to move distribution systems  towards a smarter and more secure network through integrating two-way communication infrastructure. The information exchange provides  distribution network operators (DNOs) with sophisticated management and monitoring systems to perform complex analyses and automated operations in near real-time.  Furthermore, drivers such as distribution organizations  have accelerated the  expansion of applications for smart grid technologies,  such as smart meters,  and integration of  renewable energy generators. 
 Resulting benefits  include  a more efficient use of electric appliances in households to reduce the energy bill payment for the load aggregators and lower operation cost for the generators, as well as   a higher flexibility for the DNO to enhance the system's technical operation; thereby reaching a triple-win condition. 

The DNO  is responsible for  \textit{optimal power flow} (OPF) analysis. There are challenges in  solving the OPF problem by the DNO. First, the OPF  can  be computationally difficult to be solved, especially when the number of decision variables  increases by participation of the price-responsive load aggregators in the energy market. Second, the DNO may violate the  entities' privacy, e.g., by revealing the load aggregators' demand information and generators' cost parameters to the DNO. Third, the DNO is uncertain about the load demand and  renewable generation ahead of time.

There have been some efforts in the  literature to tackle the above-mentioned challenges. We divide the related works into three main threads.  The first thread is concerned with decentralized  energy management programs for a market with multiple suppliers and multiple users. Mechnisms such as  the  multi-level game  methods \cite{Bo}, Stackelberg game \cite{maharjan}, dual decomposition method  \cite{deng}, supply bidding framework \cite{farhad}, and hierarchical bidding  \cite{ucdr4} have been used. However, these approaches did not consider the  constraints imposed by  the topology and operation of the distribution network. The second thread is concerned with including  the power flow equations in the decentralized energy management procedure. To achieve this goal, different techniques such as convex relaxation \cite{DDCOPF1,opfdr1, opfdr3, new1}, quadratic programming \cite{new2}, alternating
direction method of multipliers (ADMM) \cite{ new4, new6,new8}, and Lagrange relaxation method \cite{new5,newdr}
 have been used.  These studies, however, did not consider the uncertainties  in the  renewable generators and load demand. Furthermore, these studies mainly focused on off-line algorithms, which are applicable in day-ahead markets.  The third thread  concerned with the online operation of distribution systems  using different mechanisms such as real-time closed-loop control \cite{online1}, differential evolution optimization \cite{online3}, online gradient method \cite{online4}, projected gradient descent \cite{online5}, online mirror descent \cite{online6}, and graph theory-based approach \cite{online7}. These works, however, did not mentioned how to consider the uncertainty in the load demand  for  users in smart distribution networks.

In this paper, we  focus on designing a  distributed  algorithm  for an electricity trading market with renewable energy generators and  price-responsive  \textit{residential} load aggregators. In each time slot (e.g., every hour), the load aggregators and generators use the communication infrastructure in the smart grid to exchange information with the DNO  and jointly maximize their profit, while considering the uncertainty in the future demand and renewable  generation. The privacy of each entity is protected in the proposed  framework, as the generators and load aggregators  solve their own  profit maximization problem using the locally available information.  The main challenges that we address are  tackling the uncertainty in the load demand and renewable generation, as well as   determining the proper control signals communicated between the DNO, generators, and load aggregators that enforce the proposed distributed  algorithm to converge to the solution of the DNO's \textit{centralized} problem with the objective of maximizing the social welfare. 

The main contributions of this paper are as follows:
\begin{itemize}
\item \textit{Uncertainty Issues and Risk Evaluation}: To address the uncertainty in the load demand, we propose a probabilistic load estimation for the electric appliances of the residential users served by each load aggregator. It enables each load aggregator to schedule the electric appliances of its users in \textit{real-time}, while taking into account the  impacts of its decision  on the load profile in the upcoming time slots. We also consider an adaptive robust decision making framework for the renewable generators to optimize the risk of power shortage  based on an adjustable confidence level. It enables the generators to limit their cost for compensating the generation shortage. It also enables the DNO to prevent  high voltage drop caused by the shortage in the total renewable generation.
\item \textit{Distributed Algorithm Design}: To protect the privacy of the load aggregators and generators, as well as to address the computational complexity of the DNO's centralized problem, we propose a decentralized algorithm 
that can be executed by the entities in  real-time.  Each  entity requires to share limited information to meet its local objective, while satisfying the physical constraints of the \textit{linearized ac power flow}  in
the  distribution network.
\item \textit{Performance Evaluation}: Simulations are performed on an IEEE 123-bus test feeder with $10$ generators and $113$ load aggregators. When compared with the benchmark of not performing load management, the proposed algorithm  benefits  the load aggregators and generators by increasing their profit by $17.8\%$ and  $10.3\%$ on average, respectively. Furthermore, it helps generators to reduce the peak-to-average generation ratio by $13\%$. Our algorithm  converges to the solution of the centralized problem with a significantly  lower execution time. 
\end{itemize}

 The rest of this paper is organized as follows. Section \ref{s2} introduces the system model. Section \ref{s3} formulates the DNO's {centralized and decentralized  problems. A decentralized algorithm is proposed.}  Section IV provides  the simulation results, followed by Section \ref{s6} that concludes the paper. {Appendices A$-$F can be found in the supplementary document.}

\vspace{-4mm}

\section{System Model}\label{s2}
Consider an  electricity market with a set  $\mathcal{N}$ of $N\triangleq |\mathcal{N}|$ load aggregators and a set $\mathcal{G}$ of $G\triangleq |\mathcal{G}|$  generators scattered in a distribution network.
Each load aggregator is responsible for managing the load demand of its electricity users.  Each generator sells electricity to the market. The load aggregators and  generators use a two-way communication infrastructure  to exchange information with the DNO.  The entities are also connected to each other through the electric power distribution network. The DNO is a neutral entity  responsible for  monitoring the power flow in the network. 
  For simplicity in the problem formulation,  we assume that each bus in the network has \textit{exactly} one load aggregator or one generator. If both load aggregator  and generator  are connected to the same bus, we divide that bus into two buses connected to each other through a line with zero impedance. If neither load aggregator nor generator is connected to a bus, we propose to add a virtual load aggeragtor with zero demand for that bus. It enables us to denote the set of buses by $\mathcal{N}\cup \mathcal{G}$ and  refer a load aggregator or a generator by  its bus index. We use notation $\mathcal{L}\subseteq (\mathcal{N}\cup \mathcal{G})\times  (\mathcal{N}\cup \mathcal{G})$ to denote the set of  branches.

The overall trading horizon is denoted  by $\mathcal{H}\triangleq\{1,\dots,H\}$, where $H$ is the number of time slots with equal length (e.g., each time slot is one hour). Notice that the load management decision of a load aggregator in the \textit{current} time slot affects its demand  in the \textit{upcoming} time slots. Meanwhile, the generators need to match their generation level with the changes in the load demand. Hence, generators also need to modify their generation for the current  and upcoming time slots. To avoid an abuse of notations, hereafter, we use index $h$ for a time slot in general and use index $t$ specifically for the current time slot.

The general idea of this paper for implementing a real-time energy trading  can be summarized as follows. At the beginning of  the current time slot~$t$, the entities optimize the demand and generation profiles over the period $\mathcal{H}_t=\{t,\dots,H\}\subseteq\mathcal{H}$, but  apply only the obtained decision  for the current time slot $t$. The scheduling procedure is performed with uncertainty about the load demand and renewable generation in the upcoming time slots $h>t$. Hence, the entities repeat the optimization procedure at the beginning of  the next time slot to update their scheduling decision with the revealed demand/generation information. We aim to answer two key questions: \newline \textbf{Q.1}\, How do the entities interact with the DNO to determine their optimal load an generation profiles in the \textit{current} time slot with the locally available information? \newline \textbf{Q.2} How do the entities address the lack of information about the demand and generation in the upcoming time slots? 

\vspace{-3mm}
\subsection{Load Aggregator's Model}
In this subsection, we address questions \textbf{Q.1} and \textbf{Q.2} for residential load aggregators by modeling the  electric appliances and providing a probabilistic load estimation technique.
\subsubsection{Users' Appliances Model} Load aggregator $i\in\mathcal{N}$ is responsible for scheduling its users' electric appliances.  An electric appliance  is either \textit{\text{asleep}} or \textit{\text{awake}} in the current time slot $t$. Let $\mathcal{A}^{\text{asleep}}_i(t)$ and $\mathcal{A}^{\text{awake}}_i(t)$   denote the sets of asleep and awake appliances  in the current time slot $t$, respectively.  An awake appliance $a\in\mathcal{A}^{\text{awake}}_i(t)$ is available to be scheduled for operation, i.e., the load aggregator schedules the power consumption profile $\bm{e}_{a,i}(t) =(e_{a,i}(h),\,h\in\mathcal{H}_t)$.

The awake  appliance $a\in \mathcal{A}^{\text{awake}}_i(t)$  provides the load aggregator $i$ with  its \text{scheduling horizon}, utility function, and type using the smart meter inside the household. The scheduling horizon  $\mathcal{H}_{a,i}\subseteq \mathcal{H}_t$ defines the time interval over the upcoming time slots, in which  the appliance should be scheduled. The utility function $U_{a,i}(\bm{e}_{a,i}(t))$  is used to model the satisfaction of the customer in monetary units from using the appliance. It is generally an \textit{increasing} and \textit{concave} function of the   consumption profile $\bm{e}_{a,i}(t)$. The type of appliance depends on its specifications and the customer's  preferences. Inspired by the work in \cite{low}, we consider three types of appliances: 

$\bullet$ The appliance $a$ with \textit{type 1} should be operated  within the scheduling horizon  $\mathcal{H}_{a,i}$ and turned off  in other time slots. Examples  include the electric vehicle (EV) and dish washer. Let $\mathcal{A}_i^1(t)\subseteq \mathcal{A}^{\text{awake}}_i(t)$ denote the set of appliances with type 1 that are awake in the current time slot $t$. We have  
\begin{subequations} 
  \begin{align}
 &\!\!\!\! e_{a,i}(h)=0,\,\hspace{2.2cm} a\in\mathcal
  A_i^1(t),\, i\in\mathcal{N}, h\not\in\mathcal{H}_{a,i},\label{0}\\ 
  &\!\!\!\!e_{a,i}^{\text{min}}(h)\!\leq\! e_{a,i}(h)\!\leq \!e_{a,i}^{\text{max}}(h), \,  a\in\mathcal
  A_i^1(t),\, i\in\mathcal{N}, h\in\mathcal{H}_{a,i},\label{1}\\
&\!\!\!\!E_{a,i}^{\text{min}}\leq \textstyle\sum_{h\in\mathcal{H}_{a,i}}e_{a,i}(h)\leq E_{a,i}^{\text{max}},\hspace{0.4cm} a\in\mathcal
  A_i^1(t),\, i\in\mathcal{N}.\label{2}
\end{align}
\end{subequations}
 The utility  obtained from using a  type 1 appliance depends on the total power consumption.  The  utility can be expressed as $U_{a,i}(\bm{e}_{a,i}(t))=U_{a,i}\big(\textstyle\sum_{h\in\mathcal{H}_{a,i}}e_{a,i}(h)\big)$, e.g., $U_{a,i}\big(\textstyle\sum_{h\in\mathcal{H}_{a,i}}e_{a,i}(h)\big)=\kappa_{a,i} f\big(\sum_{h\in\mathcal{H}_{a,i}}e_{a,i}(h)-E_{a,i}^{\text{min}}\big)$ with a concave function $f(\cdot)$  and nonnegative constant $\kappa_{a,i}$.

$\bullet$ The appliances of \textit{type 2}  can be operated in  time slots out of the scheduling horizon, but the customer  attains  a relatively low utility, e.g., TV and personal computer. Let $\mathcal{A}_i^2(t)\subseteq \mathcal{A}^{\text{awake}}_i(t)$ denote the set of appliances of type 2 that are awake in the current time slot $t$. We have
\begin{subequations} 
  \begin{align}
 & \!\!\!e_{a,i}(h)\geq 0,\,\hspace{2.1cm} a\in\mathcal
  A_i^2(t),\, i\in\mathcal{N}, h\not\in\mathcal{H}_{a,i},\label{4}\\ 
  &\!\!\!e_{a,i}^{\text{min}}(h)\!\leq\! e_{a,i}(h)\!\leq\! e_{a,i}^{\text{max}}(h),  a\in\mathcal
  A_i^2(t),\, i\in\mathcal{N}, h\in\mathcal{H}_{a,i},\label{5}\\
&\!\!\!E_{a,i}^{\text{min}}\leq \textstyle\sum_{h\in\mathcal{H}_{a,i}}e_{a,i}(h)\leq E_{a,i}^{\text{max}},\hspace{0.3cm} a\in\mathcal
  A_i^2(t),\, i\in\mathcal{N}.\label{6}
\end{align}
\end{subequations}
The utility function for type 2 appliances depends on both the amount of power consumption  and the time of consuming the power, i.e., the customer would gain different benefits from consuming the same amount of power at different times, e.g., watching the favorite TV program. We have $U_{a,i}(\bm{e}_{a,i}(t))=\sum_{h\in\mathcal{H}_t}U_{a,i}(e_{a,i}(h),h)$. As a concrete example, utility function $U_{a,i}(\bm{e}_{a,i})=\sum_{h\in\mathcal{H}_{a,i}}\kappa_{a,i}(h) f\big(e_{a,i}(h)-e_{a,i}^{\text{min}}\big)+\sum_{k\not\in\mathcal{H}_{a,i}}\kappa'_{a,i}(h) f\big(e_{a,i}(h)\big)$ with a concave function $f(\cdot)$ and  \textit{time dependent} nonnegative coefficients $\kappa_{a,i}(h)$ and $\kappa'_{a,i}(h)$, $\kappa'_{a,i}(h)\ll\kappa_{a,i}(h)$ is a viable candidate.

$\bullet$ The appliances of \textit{type 3}  can  be operated  out of the scheduling horizon  without any constraint on their total power consumption, such as lighting and refrigerator. Let $\mathcal{A}_i^3(t)\subseteq \mathcal{A}^{\text{awake}}_i(t)$ denote the set of appliances of type 3 that are awake in the current time slot $t$. We have
\begin{subequations} 
  \begin{align}
 &\!\!\!\! e_{a,i}(h)\geq 0,\,\hspace{2.19cm} a\in\mathcal
  A_i^3(t),\, i\in\mathcal{N}, h\not\in\mathcal{H}_{a,i},\label{20}\\ 
  &\!\!\!\!e_{a,i}^{\text{min}}(h)\!\leq\! e_{a,i}(h)\!\leq \!e_{a,i}^{\text{max}}(h), \,  a\in\mathcal
  A_i^3(t),\, i\in\mathcal{N}, h\in\mathcal{H}_{a,i}.\label{21}
\end{align}
\end{subequations}
The utility $U_{a,i}(\bm{e}_{a,i})$ attained by the customer from using the appliances with type 3 depends on the amount of power consumption $e_{a,i}(t)$ within the scheduling horizon $\mathcal{H}_{a,i}$, but not the time of consumption.   The customer  attains  a relatively low utility out of interval $\mathcal{H}_{a,i}$.   Function $U_{a,i}(\bm{e}_{a,i})=\sum_{h\in\mathcal{H}_{a,i}}\kappa_{a,i} f\big(e_{a,i}(h)-e_{a,i}^{\text{min}}\big)+\sum_{h\not\in\mathcal{H}_{a,i}}\kappa'_{a,i} f\big(e_{a,i}(h)\big)$ with a concave function $f(\cdot)$ and nonnegative \textit{constants} $\kappa_{a,i}$ and $\kappa'_{a,i}$, $\kappa'_{a,i}\ll\kappa_{a,i}$ is a viable candidate.

The total utility of load aggregator $i$ in the current time slot $t$ with decision vector $\bm{e}_i(t)=(\bm{e}_{a,i}(t),\,a\in\mathcal{A}^{\text{awake}}_i(t))$ is
\begin{align}\label{totalutil}
U_i(\bm{e}_i(t))=\textstyle\sum_{a\in\mathcal{A}^{\text{awake}}_i(t)}\!U_{a,i}(\bm{e}_{a,i}(t)),\;\;\;\; i\in\mathcal{N}.
\end{align} 

\subsubsection{Load Estimation} The  actual wake-up times of the appliances are not available to the load aggregator in advance. To address this lack of information,  load aggregator $i$ can collect the sleep-awake historical data record of each appliance and estimate the probability $p_{a,i}(h)$ that each appliance $a\in\mathcal{A}_i$ becomes awake at each time slot $h\in\mathcal{H}$.  In appendix A, we show   the \textit{conditional} probability $p_{a,i}(h\,|\,t)$ that the appliance becomes awake in an upcoming time slot $h > t$, given that it has not become
awake until the current time slot, $t$, is
\begin{align}\label{prob}
p_{a,i}(h\,|\,t)=\frac{p_{a,i}(h)}{1-\sum_{h'=1}^t p_{a,i}(h')}.
\end{align}

A load aggregator has no information about the scheduling horizon, users' utility, and  type of the appliances ahead of time. For decision making at the current time slot $t$,   we consider the \textit{worst-case scenario}, in which the electric appliances that become awake in the upcoming time slots $h>t$ should be operated once they become awake without any control on power consumption, i.e., $e_{a,i}^{\text{min}}(h)=  e_{a,i}^{\text{max}}(h)=e_{a,i}^{\text{nom}}$ and  $E_{a,i}^{\text{min}}=  E_{a,i}^{\text{max}}=E^{\text{nom}}_{a,i}$.  The   payment  of the load aggregator in the worst-case scenario is an upper-bound for its actual payment. Hence, minimizing the worst-case payment implies reducing the risk of high  payment. For the current time slot $t$,  the worst-case expected electric demand $l^{\text{asleep}}_i(h)$ of the currently sleeping appliances in an upcoming time slot $h>t$~is
\begin{align}\label{predict}
\!\!\!\!l^{\text{asleep}}_i(h)= &\sum_{a\in\mathcal{A}^{\text{asleep}}_i(t)}\!\!\!\!\!\!e_{a,i}^{\text{nom}}(h)\Bigg[\sum_{h'=\max\{t+1,h-T_a+1\}}^h\!\!\!\!\!\!\!\!\!\! p_{a,i}(h'\,|\,t)\Bigg].
\end{align}
\noindent where  parameter $T_a=E^{\text{nom}}_{a,i}/e_{a,i}^{\text{nom}}$ is the operation duration of the appliance $a\in\mathcal{A}^{\text{asleep}}_i(t)$ that becomes awake in
upcoming time slot $h>t$. The value of $\sum_{h'=\max\{t+1,h-T_a+1\}}^k p_{a}(h'\,|\,t)$ is equal to the probability
that a currently  sleeping appliance $a$ is operating in the upcoming time slot $h>t$. 

For the given current time slot $t$, we use the notation $\bm{l}_i(t)=(l_i(h),\,h\in\mathcal{H}_t)$ to denote the profile of \textit{active power}  consumption of the users during  time interval $\mathcal{H}_t$. We have
\begin{align}\label{predict2}
l_i(h)= &l^{\text{asleep}}_i(h)+\textstyle\sum_{a\in\mathcal{A}^{\text{awake}}_i(t)}e_{a,i}(h),\;\;\;\;\;h\in\mathcal{H}_t.
\end{align}

To model the \text{reactive power} consumption for a load aggregator $i$, we consider a constant power factor $\text{PF}_i$. The reactive power for load aggregator $i$ in time $h$ is $q_i(h)=l_i(h)\sqrt{\frac{1-\text{PF}_i^2}{\text{PF}_i^2}}$.

\subsubsection{Local Optimization Problem}  Constraints (\ref{0})$-$(\ref{predict2}) define the feasible  set $\mathcal{E}_i(t)$ for decision vector $\bm{e}_i(t)$ of load aggregator $i$ in the current time slot $t$.
Load aggregator $i$ aims to maximize the profit  $\pi^{\text{agg}}_i(\boldsymbol{e}_i(t))$, which includes  the  total utility in (\ref{totalutil}) minus the  payment to the DNO over  period $\mathcal{H}_t$. The DNO provides load aggregator $i$ with the price $\rho_{i}(h)$ for a unit of active power in each time slot $h$. We assume that load aggregators do not pay for the reactive power. We have
\begin{align}\label{costt}
\pi^{\text{agg}}_i(\boldsymbol{e}_i(t))=U_i(\bm{e}_i(t))-\textstyle\sum_{h\in\mathcal{H}_t} l_i(h)\,\rho_i(h), \;\;\;i\in\mathcal{N},
\end{align}

 Load aggregator $i$ solves the following optimization problem in time slot $t$ to determine decision vector $\bm{e}_i(t)$:
 \begin{subequations}\label{agg_prob} 
\begin{align}
	&\underset{\boldsymbol{e}_i(t)}{\text{maximize}}\;\; \pi^{\text{agg}}_i(\boldsymbol{e}_i(t)) & 	\\
&\text{subject to} \;\; \bm{e}_i(t)\in\mathcal{E}_i(t). \hspace{0cm}\label{demand}
\end{align}
\end{subequations}

\vspace{-6mm}
\subsection{Generator's Model}
In this subsection, we address questions Q.1 and Q.2 for the generators by modeling the conventional and renewable units and providing a robust optimization technique for renewables.
\subsubsection{Conventional Unit} In general, the generation cost function of the conventional unit of generator $j\in\mathcal{G}$ in time slot $h$ is  an increasing convex function of  $p^{\text{conv}}_{j}(h)$ \cite{genfunc}.  The class of quadratic  functions $C_{j}(p^{\text{conv}}_{j}(h))=a_{j2}\left(p^{\text{conv}}_{j}(h)\right)^2$ $+a_{j1}p^{\text{conv}}_{j}(h)+a_{j0}$ is well-known.
 For the given current time slot $t$, generator $j$ offers the profiles of active and reactive powers  $\bm{p}^{\text{conv}}_j(t)=(p^{\text{conv}}_{j}(h),\,h\in\mathcal{H}_t)$ and $\bm{q}^{\text{conv}}_j(t)=(q^{\text{conv}}_{j}(h),\,h\in\mathcal{H}_t)$  for the current and upcoming time slots.

\subsubsection{Renewable Unit}   Without loss of generality, we assume that the renewable plants are operated at unity power factor.
 Given the current time slot $t$, generator $j$ with renewable unit offers an active power profile $\bm{p}^{\text{ren}}_j(t)=(p^{\text{ren}}_j(h),\,h\in\mathcal{H}_t)$ for its renewable unit. To prevent non-credible high renewable generation offers, the DNO charges generator $j$ by the unit price $\beta_{j}(h)$ (\$/MW) for generation shortage in time slot $h$. To cope with the uncertainty issues, we consider a \textit{robust} decision making for generators with renewable units. Generator $j$ can uses the historical data record to forecast an uncertainty bound $[p^{\text{min,ren}}_j(h),p^{\text{max,ren}}_j(h)]$ for its \textit{actual} renewable generation in  time slot $h\in\mathcal{H}$. Generator $j$ considers the cost $\Gamma_{j}(\bm{p}^{\text{ren}}_j(t))$ of the worst-case scenario for generation shortage as follows:
\begin{align}\label{gamma22}
&\Gamma_{j}(\bm{p}^{\text{ren}}_j(t))\triangleq \textstyle{\sum_{h\in\mathcal{H}_t}}\, \beta_{j}(h)(p^{\text{ren}}_j(h)-p^{\text{min,ren}}_j(h)).
\end{align}
The feasible set for the renewable generation profile $\bm{p}^{\text{ren}}_j(t)$ can be defined based on all scenarios that satisfy  $p^{\text{ren}}_j(h)\in[p^{\text{min,ren}}_j(h),p^{\text{max,ren}}_j(h)]$. However, it is very conservative and possibly inefficient to take into account all possible scenarios. Inspired by the work in \cite{adaptive}, we consider an adaptive robust model. In the  current time slot $t$, the  uncertainty space for the  generation profile $\bm{p}^{\text{ren}}_j(t)$ in the time interval $\mathcal{H}_t$ is defined~as
\begin{align} \label{uncertainty2}
\!\!\!\!{\mathcal{P}}^{\text{ren}}_j(t)=&\Big\{{\bm{p}}^{\text{ren}}_j(t)\,|\,{p}^{\text{ren}}_j(h)\in[p^{\text{min,ren}}_j(h),p^{\text{max,ren}}_j(h)],\, h\in\mathcal{H}_t,\nonumber\\& \;\;\;\textstyle{\sum_{h\in\mathcal{H}_t} } \dfrac{p^{\text{max,ren}}_j(h)-{p}^{\text{ren}}_j(h)}{p^{\text{max,ren}}_j(h)-p^{\text{min,ren}}_j(h)}\leq\Delta_j(t)\Big\},
\end{align}
where $0\leq\Delta_j(t)\leq |\mathcal{H}_t|$ is the confidence level parameter for generator $j$ in the current time slot $t$. The space defined in  (\ref{uncertainty2}) is a singleton, corresponding to the least-conservative scenario ${p}^{\text{ren}}_j(h)=p^{\text{max,ren}}_j(h),\,h\in\mathcal{H}_t$ when $\Delta_j(t)=0$. As $\Delta_j(t)$ increases, the size of the
uncertainty set enlarges, and the
resulting robust solution is more conservative. The space includes all possible scenarios  when $\Delta_j(t)=|\mathcal{H}_t|$. In  \cite{adaptive}, $\Delta_j(t)$ is known and fixed. Whereas,  we consider parameter $\Delta_j(t)$ as a variable that should be optimized by generator $j$.

\subsubsection{Local Optimization Problem} For a given current time slot $t$,  generator $j$ decides on the generation profile $\bm{\psi}_j(t)=(\bm{p}^{\text{conv}}_j(t), \bm{q}^{\text{conv}}_j(t), \bm{p}^{\text{ren}}_j(t))$ and the confidence level $\Delta_j(t)$.  
   The  objective of   generator $j$ is to maximize its profit $\pi^{\text{gen}}_j(\bm{\psi}_j(t))$, which is    the revenue from selling active and reactive powers minus the generation cost and  financial risk   in (\ref{gamma22}). That is
\begin{align}\label{supprob}
\!\!\pi^{\text{gen}}_j&(\bm{\psi}_j(t))=\textstyle\sum_{h\in\mathcal{H}_t}\big[\big(p^{\text{conv}}_j(h)\!+\!p^{\text{ren}}_j(h)\big)\rho_{j}(h)\nonumber\\&+q^{\text{conv}}_j(h)\varrho_{j}(h)-C_{j}(p^{\text{conv}}_{j}(h))\big]-\Gamma_{j}(\bm{p}^{\text{ren}}_j(t)).
\end{align}
 The problem for generator $j\in\mathcal{G}$ in the current time slot $t$ is 
 \vspace{-1mm}
 \begin{subequations}\label{util_problem}
 \begin{align}\label{util_obj}
&\underset{\bm{\psi}_j(t),\,\Delta_j(t)} {\text{maximize}}\,\;\pi^{\text{gen}}_j(\bm{\psi}_j(t))\\
&\text{subject to}\;\;p^{\text{min,conv}}_{j}\leq p^{\text{conv}}_{j}(h)\leq p^{\text{max,conv}}_{j},\hspace{0.2cm}  h\in\mathcal{H}_t,\label{gen11}\\
&\hspace{1.6cm}q^{\text{min,conv}}_{j}\leq q^{\text{conv}}_{j}(h)\leq q^{\text{max,conv}}_{j},\hspace{0.2cm}  h\in\mathcal{H}_t,\label{gen2}\\
&\hspace{1.6cm}{\bm{p}}^{\text{ren}}_j(t)\in\mathcal{P}^{\text{ren}}_j(t).\label{gen3}
 \end{align}
\end{subequations}

   \vspace{-6mm}  
\subsection{DNO's Model}
\vspace{-1mm}
We address  Q.1 and Q.2 for the DNO in the following.

\vspace{-0mm}
\subsubsection{Linearized ac Power flow} We consider balanced distribution networks. The model for unbalanced networks is a research for future work. The ac power flow equations are nonlinear and nonconvex. An  alternative is to use a linearized  ac power flow.  Let $\bm{p}(h)=(p_b(h),\, b\in\mathcal{N}\cup\mathcal{G})$ and $\bm{q}(h)=(q_b(h),\, b\in\mathcal{N}\cup\mathcal{G})$ denote the vectors of injected active power $p_b(h)$ and reactive power  $q_b(h)$ to all buses $b\in\mathcal{N}\cup\mathcal{G}$ in time  $h$. Let  $|v_{b}(h)|$ and $\theta_{b}(h)$  denote the voltage magnitude and phase angle of   bus $b$ in time $h$. We define the grid-wide vectors $\bm {\theta}(h)=\left(\theta_{b}(h),\, b\in\mathcal{N}\cup\mathcal{G}\right)$ and  $|\bm{v}(h)|=\left(|v_{b}(h)|,\, b\in\mathcal{N}\cup\mathcal{G}\right)$  in time slot $h$. Let $G_{rs}$ and $B_{rs}$ denote the real and reactive parts of the
entry  $(r,s)$ in bus admittance matrix $Y$. Let $b_{rr}$ and $g_{rr}$ denote the  shunt susceptance and conductance at bus $r$. In Appendix~B, we show that the linearized ac power flow  can be expressed~as
\begin{equation}\label{acpf}
  \begin{bmatrix}
    \bm{p}(h) \\
 \bm{q}(h)
  \end{bmatrix}=\underbrace{\begin{bmatrix}
    -\bm {B}'&\;\;\bm {G}'\\
-\bm {G}'&-\bm {B}
  \end{bmatrix}}_{\bm{\Lambda}}\begin{bmatrix}
    \bm {\theta}(h)\\
|\bm{v}(h)|
  \end{bmatrix},
\end{equation}
where the diagonal element $(r,r)$ of matrices $\bm {B}$ and $\bm {B}'$ are $B_{rr}$ and $B_{rr}-b_{rr}$, respectively. The non-diagonal elements $(r,s)$ of both $\bm {B}$ and $\bm {B}'$ are $B_{rs}$. The diagonal element $(r,r)$ of matrices $\bm {G}$ and $\bm {G}'$ are $G_{rr}$ and $G_{rr}-g_{rr}$, respectively. The non-diagonal elements $(r,s)$ of both $\bm {G}$ and $\bm {G}'$ are $G_{rs}$. 
In Appendix C, we show that, in time slot $h$, the linearized active and reactive power flow  through  line $(r,s)\in\mathcal{L}$ with resistance $R_{rs}$ and reactance $X_{rs}$ can be calculated as:
\begin{subequations}\label{flow}
\begin{align}
&\!\!\!p_{rs}(h)= \frac{R_{rs}(|v_r(h)|\!-\!|v_s(h)|)\!+\!X_{rs}(\theta_r(h)\!-\!\theta_s(h))}{R_{rs}^2+X_{rs}^2},\\
&\!\!\!q_{rs}(h)= \frac{X_{rs}(|v_r(h)|\!-\!|v_s(h)|)\!-\!R_{rs}(\theta_r(h)\!-\!\theta_s(h))}{R_{rs}^2+X_{rs}^2},
\end{align}
\end{subequations}

The apparent power flow $s_{rs}(h)=\sqrt{p_{rs}^2(h)+q^2_{rs}(h)}$ is upper bounded by  $s_{rs}^{\text{max}}$, which implies that the feasible real and reactive powers  are bounded by a circle. To linearize the constraint, a piecewise approximation of the boundary by a regular polygon with central angle $\alpha$ can be used. 
In Appendix~D, we obtain the following  constraints
\begin{align}\label{apparent}
& p_{rs}(h)\,\text{cos}\, (m\alpha) + q_{rs}(h)\,\text{sin}\, (m\alpha)\,\leq s_{rs}^{\text{max}},
\end{align}
where $m=0,\dots,2\pi/\alpha$. For each bus $b$, we also have
\begin{align}\label{voltage}
v_b^{\text{min}}\leq |v_b(h)|\leq v_b^{\text{max}}.
\end{align}

\vspace{-1mm}
\subsubsection{DNO's Centralized Optimization Problem} The DNO  considers the impact of renewable generation shortage  on the technical operation of the network. As a concrete example, the risk of voltage drop at different buses is a viable choice for the DNO. The DNO  considers   function $\Gamma^{\text{DNO}}(t)$ of the grid-wide renewable generation $\bm{p}^{\text{ren}}(t)$ for the voltage variations as
\begin{align}\label{gamma2}
&\!\!\!\!\Gamma^{\text{DNO}}(\bm{p}^{\text{ren}}(t))\triangleq \textstyle{\sum_{h\in\mathcal{H}_t} \; \sum_{b\in\mathcal{N}\cup\mathcal{G}}}\left(|v_b(h)|-|\widehat{v}_b(h)|\right),
\end{align}
where $\widehat{v}_b(h)$ is voltage magnitude of bus $b$ in the worst-case scenario, when all renewable generators' power are $p^{\text{min,ren}}_j(h)$ in time slot $h\in\mathcal{H}_t$. The uncertainty space that the DNO considers for renewable generator $j$ is defined as (\ref{uncertainty2}).

We consider the objective of maximizing the social welfare for the DNO. 
 Considering the grid-wide vectors $\bm{e}(t)$ and $\bm{p}^{\text{conv}}(t)$, and $\bm{p}^{\text{ren}}(t)$, the DNO's objective function is
\begin{align}\label{EPAR}
f^{\text{DNO}}&\left(\bm{e}(t),\bm{p}^{\text{conv}}(t),\bm{p}^{\text{ren}}(t)\right)\triangleq \,\textstyle\sum_{i\in\mathcal{N}}\, U_i(\bm{e}_i(t))\!\nonumber\\&-\textstyle\sum_{h\in\mathcal{H}_t}\sum_{j\in\mathcal{G}}\! C_{j}(p^{\text{conv}}_{j}\!(h))-\!\vartheta^{\text{c}}\,\Gamma^{\text{DNO}}(\bm{p}^{\text{ren}}(t)),
\end{align}

\noindent where $\vartheta^{\text{c}}$  is a positive weighting coefficient.
 We formulate the DNO's \textit{centralized  problem} as
 \begin{subequations}  \label{ISO_problem}
\begin{align}\label{ISO_obj}
&\hspace{-1.7cm}\underset{\begin{subarray}{c}
\hspace{1cm}\bm{e}(t),\bm{p}^{\text{conv}}(t),\bm{q}^{\text{conv}}(t),\\\hspace{1.4cm}\bm{p}^{\text{ren}}(t),\bm{\Delta}(t),|\bm{v}(t)|,\bm{\theta}(t)
\end{subarray}} {\text{maximize}}\,\,\,f^{\text{DNO}}\left(\bm{e}(t),\bm{p}^{\text{conv}}(t),\bm{p}^{\text{ren}}(t)\right)\\
&\hspace{-0.3cm}\text{subject to constraints (\ref{demand}),\,(\ref{gen11})$-$(\ref{gen3}), (\ref{acpf})$-$(\ref{gamma2})}.\label{ISO0}
\end{align}
\end{subequations}
 Problem (\ref{ISO_problem}) has a concave  objective function and linear  constraints due to the concavity of the load aggregators' utility function, the convexity of the generation cost function, and the linearity of the ac power flow model in (\ref{acpf}). {\color{black}Hence, we have:
}
\vspace{0.0mm}
\noindent\textbf{Theorem 1} \textit{The optimal solution to the  DNO's centralized problem in (\ref{ISO_problem}) is unique.}
\vspace{0.0mm}

To solve  the centralized problem  (\ref{ISO_problem}), the DNO needs  the information about the load aggregators' utilities, generators' generation cost, and renewable units'  forecast data. However, these information may not be available to the DNO. 
Instead, we develop a decentralized algorithm by showing that the DNO can determine  $\rho_b(h)$ and $\varrho_b(h),\,b\in\mathcal{N}\cup\mathcal{M}$, as well as the penalties $\beta_j(h),\,j\in\mathcal{G}$ for  $h\in\mathcal{H}_t$ such that when the load aggregators solve (\ref{agg_prob}) and generators solve  (\ref{util_problem}), the resulting solution coincides with the \text{unique solution} of  problem  (\ref{ISO_problem}).

\vspace{-2.5mm}
\section{Decentralized Algorithm Design}\label{s3}

 Given the current time slot $t$, the decision vector of  load aggregator $i$ is load profile  $\boldsymbol{e}_i(t)$ of the  awake appliances. Further, the decision vector of generator $j$ is the generation profile $\bm{\psi}_j(t)=(\bm{p}^{\text{conv}}_j(t), \bm{q}^{\text{conv}}_j(t), \bm{p}^{\text{ren}}_j(t))$ and the confidence level $\Delta_j(t)$.   The DNO   influences 
the entities by using the nodal prices $\boldsymbol{\rho}(t)$, $\boldsymbol{\varrho}(t)$, and penalties $\bm{\beta}(t)$.  

One well-know technique to determine the proper values of the nodal prices $\boldsymbol{\rho}(t)$, $\boldsymbol{\varrho}(t)$, and penalties $\bm{\beta}(t)$ is to formulate the partial \textit{Lagrangian relaxation} of the DNO's centralized problem (\ref{ISO_problem}) \cite{new5,new7,newdr}. Let $\lambda_b(h)$ and $\gamma_b(h),\,b\in\mathcal{N}\cup\mathcal{G},\, h\in\mathcal{H}$ denote the Lagrange multipliers associated with the equality constraints for the injected active power $p_b(h)$ and reactive power $q_b(h)$ in (\ref{acpf}). We move these constraints with their Lagrange multipliers to the objective function of the  centralized problem in (\ref{ISO_problem}). In equation (S-12) of Appendix E, we obtain the objective function $f_{\text{Lag}}^{\text{DNO}}(\cdot)$ of the relaxed problem. Due to the convexity of problem (\ref{ISO_problem}) and linearity of the constraints, the strong duality gap condition (Slater's condition) is satisfied if a feasible solution exist \cite[Ch. 5]{boyd2004convex}. Thus, the optimal solution to the  relaxed problem is equal to the optimal solution to the primal problem (\ref{ISO_problem}).  Using the relaxed problem enables us to determine the price signals  $\boldsymbol{\rho}(t)$, $\boldsymbol{\varrho}(t)$, and $\boldsymbol{\beta}(t)$ based on the \textit{Lagrangian decomposition} technique, such that  the market equilibrium among load aggregators and generators coincides with the optimal solution of the centralized problem (\ref{ISO_problem}). 
 
 \vspace{0.0mm}
\noindent\textbf{Theorem 2} \textit{The equilibrium of the energy market coincides with the unique solution to the  DNO's centralized problem in (\ref{ISO_problem}) if and only  if for $i\in\mathcal{N},j\in\mathcal{G},\, h\in\mathcal{H}_t$ the DNO sets }
\begin{subequations}
\begin{align}
&\!\!\!\rho_i(h)=\lambda_i(h)+\gamma_i(h)\sqrt{\sfrac{(1-\text{PF}_i^2)}{\text{PF}_i^2}},\hspace{0.2cm}i\in\mathcal{N},\, h\in\mathcal{H}_t,\label{penal1}\\
&\!\!\!\rho_j(h)=\lambda_j(h),\hspace{3.4cm}j\in\mathcal{G},\,\, h\in\mathcal{H}_t,\label{penal2}\\
&\!\!\!\varrho_j(h)=\gamma_j(h),\hspace{3.4cm}j\in\mathcal{G},\,\, h\in\mathcal{H}_t,\label{penal3}\\
& \!\!\!\beta_j(h)=\vartheta^{\text{c}}\textstyle\sum_{b\in\mathcal{N}\cup\mathcal{G}}\bm{\Lambda}^{-1}(|\mathcal{N}\cup\mathcal{G}|+b,j),\label{penal4}
\end{align}
\end{subequations}
\textit{where $\bm{\Lambda}^{-1}(|\mathcal{N}\cup\mathcal{G}|+b,j)$ is the entry $(|\mathcal{N}\cup\mathcal{G}|+b,j)$ of the inverse of matrix $\bm{\Lambda}$ in (\ref{acpf})}.
%

\noindent The proof can be found in Appendix E.  It suggests a decentralized algorithm to determine the solution to problem (\ref{ISO_problem}).

 We propose Algorithm 1 that can be executed by the load aggregators, generators, and DNO in real-time. In Algorithm 1, when the current time slot $t$ begins, each load aggregator $i$ determines the power consumption profile $\bm{e}_{a,i}(t) =(e_{a,i}(h),\,h\in\mathcal{H}_t)$ of all awake appliances $a\in\mathcal{A}^{\text{awake}}_i(t)$ over  time slots $h\in\mathcal{H}_t$.    Each generator $j$ obtains the profiles of active and reactive powers  $\bm{p}^{\text{conv}}_j(t)=(p^{\text{conv}}_{j}(h),\,h\in\mathcal{H}_t)$ and $\bm{q}^{\text{conv}}_j(t)=(q^{\text{conv}}_{j}(h),\,h\in\mathcal{H}_t)$ of the conventional unit and generation profile $\bm{p}^{\text{ren}}_j(t)=(p^{\text{ren}}_{j}(h),\,h\in\mathcal{H}_t)$.  The entities use the obtained scheduling decision for upcoming time slots $h\geq t+1$ as an \textit{initial}  decision in the next time slot $t+1$.  

%

 In each time slot $t$, Algorithm 1 is executed in an iterative fashion.    Let $k$ denote the iteration index.   Our algorithm involves the {initiation phase} and { market trading phase}.

\noindent\noindent$\bullet$\;\textit{Initiation phase}: Lines 1 to 9 describe the initiation phase. 

\noindent$\bullet$\,\textit{Market trading phase}:  The loop involving  Lines 10 to 18 describes this phase, which includes the following parts:

 \textit{ a) Information exchange}:  In Line 11, each load aggregator~$i$ uses (\ref{predict2}) to obtain its  demand profile $\bm{l}_i^k(t)=(l^k_i(h), h\in\mathcal{H}_t)$, and sends    to the DNO. Each generator~$j$  sends the  profiles $\bm{p}_{j}^{\text{conv},k}(t)$, $\bm{q}^{\text{conv},k}_{j}(t)$, and $\bm{p}_{j}^{\text{ren},k}(t)$ to the~DNO. \vspace{0.0mm}

\textit{b) DNO's update}: In Line 12, the DNO receives the information from the entities, it obtains the updated vector $\bm{\phi}^{k+1}\!(t)\!=\!(\bm{\theta}^{k+1}_{b}(t),|\bm{v}^{k+1}_b(t)|, \bm{\lambda}^{k+1}_b(t),\!\bm{\gamma}^{k+1}_b(t),b\in\mathcal{N}\!\cup\mathcal{G})$~as 
\begin{align}
     	\bm{\phi}^{k+1}(t)=\left[\boldsymbol{\phi}^{k}(t)+\epsilon^k\nabla_{\bm{\phi}^{k}(t)}f_{\text{Lag}}^{\text{DNO}}\!\left(\cdot\right)\right]^{+},  \label{signal2}
     	   \end{align}

\noindent where $\nabla$ is the gradient operator, and $[\cdot]^{+}$ is the projection onto the feasible set defined by constraints in (\ref{ISO0}). Recall that $f_{\text{Lag}}^{\text{DNO}}\!\left(\cdot\right)$ is the objective function of the DNO's relaxed problem, which is given in equation (S-12) in Appendix  E. The DNO  uses (\ref{penal1})$-$(\ref{penal4})  to compute the updated   prices $\bm{\rho}^{k+1}(t)$ and $\bm{\varrho}^{k+1}(t)$ and penalties $\bm{\beta}^{k+1}(t)$ for all buses.

\setlength{\textfloatsep}{4pt}
\begin{algorithm}[t]\label{al1}\footnotesize
 	\caption{ Decentralized Energy Market Trading Algorithm.}
 	\begin{algorithmic} [1]
 	
 	\STATE  Set $k:= 1$ and $\xi_1=\xi_2:=10^{-2}$.
 	\STATE \textbf{If} $t=1$
 	\STATE Each load aggregator  $i\in\mathcal{N}$ randomly initializes its users' appliances load profile $\boldsymbol{e}^{1}_i(t)$. 
 	\STATE Each generator $j\in\mathcal{G}$ randomly initializes its conventional generation profiles $\boldsymbol{p}_j^{\text{conv},1}(t)$ and $\boldsymbol{q}_j^{\text{conv},1}(t)$.  
 	\STATE Each generator $j$ with renewable units  initializes $\Delta_j^1(t)=0$ and set the presumed generation levels to $p_j^{\text{ren},1}(h)=p_j^{\text{max,ren}}(h)$ for $h\in\mathcal{H}_t$.
 	\STATE The DNO   sets $|v_b^1(h)|=1$ pu,  $\theta_b^1=0,$ and $\lambda^1_b(h)=\gamma^1_b(h)=0,\,b\in\mathcal{N}\cup\mathcal{G}, \,h\in\mathcal{H}_t$.
 	\STATE \textbf{Else if} $t>1$
 	\STATE Load aggregators, generators, and DNO initialize their decision variables with their values in the equilibrium at previous time slot $t-1$.
 	\STATE\textbf{End if}

\STATE \textbf{Repeat}
\STATE \hspace{1mm} Each load aggregator  $i$ and generator $j$ sends its  load  profile $\bm{l}_i^k(t)$ and \\ \hspace{1mm}  generation profiles \,$\bm{p}_{j}^{\text{conv},k}(t)$, $\bm{q}^{\text{conv},k}_{j}(t)$,    and $\bm{p}_{j}^{\text{ren},k}(t)$  to the DNO.
\STATE \hspace{1mm} DNO obtains the updated vector $\bm{\phi}^{k+1}(t)=(\bm{\theta}^{k+1}_{b}(t),|\bm{v}^{k+1}_b(t)|, $ \\ \hspace{1mm} $\bm{\lambda}^{k+1}_b(t),\bm{\gamma}^{k+1}_b(t),\,b\in\mathcal{N}\cup\mathcal{G})$ using (\ref{signal2}).
 	
 	\STATE \hspace{1mm}  DNO  uses (\ref{penal1})$-$(\ref{penal4})  to compute the updated values of control signals \\ \hspace{1mm} $\bm{\rho}^{k+1}(t)$,   $\bm{\varrho}^{k+1}(t)$, and  $\bm{\beta}^{k+1}(t)$, and sends the control signals  to the \\ \hspace{1mm} corresponding entity in each bus.
 	\STATE \hspace{1mm} Each load aggregator $i$ updates  its controllable load profile $\boldsymbol{e}^{k+1}_i(t)$ by solving its local problem (\ref{agg_prob})
 	\STATE \hspace{1mm} Each generator $j$  updates its  generation profile $\bm{\psi}_j^{k}(t)$ and  decision \\ \hspace{1mm} variable $\Delta_j^{k}(t)$, by solving its local problem  in (\ref{util_problem}).
 	 	\STATE \hspace{1mm} $k:=k+1$. The step size  is updated.
 	
 	\STATE \textbf{Until}   $|\bm{\theta}_{b}^{k}(t)-\bm{\theta}_{b}^{k-1}(t)|\leq \xi_1, ||\bm{v}_{b}^{k}(t)|-|\bm{v}_{b}^{k-1}(t)||\leq \xi_2,\,b\in\mathcal{N}\cup\mathcal{G}$.
 	 	\end{algorithmic} 
 \end{algorithm} 
 \normalsize

 {\color{black}\textit{ c) Load aggregator's update}:  When load aggregator  $i$ receives the control signal $\rho^{k+1}_{i}(h),\,h\in\mathcal{H}_t$ from the DNO, in Line 9, it updates its controllable load profile $\boldsymbol{e}^{k+1}_i(t)$ by solving its local problem (\ref{agg_prob}), which is  convex  and can be efficiently solved at each iteration.  Note that the utility function in (\ref{costt}) is a concave function.


 \textit{d) Generator's update}:  When generator $j$ receives  signals $\bm{\rho}^{k+1}_j(t)$, $\bm{\varrho}^{k+1}_j(t)$ and $\bm{\beta}^{k+1}_j(t)$, it updates its  generation profile $\bm{\psi}_j^{k}(t)=(\bm{p}^{\text{conv},k}_j(t), \bm{q}^{\text{conv},k}_j(t),$ $ \bm{p}^{\text{ren},k}_j(t))$ and  decision variable $\Delta_j^{k}(t)$, by solving its local problem  in (\ref{util_problem}). This problem is a linear problem and
 can be solved efficiently by the generator using its local
 information about its conventional and renewable units.

     	   
     	 }

 \textit{e) Step size update:} We use a nonsummable diminishing step size.   In Line 17, the step size is updated. 
 
  We emphasize that in Algorithm 1, the DNO needs to consider on bus (e.g., the substation bus) as the slack bus.




\vspace{-1.5mm}
\section{Performance Evaluation}\label{s5}
\vspace{-1mm}
In this section, we evaluate the performance of our proposed decentralized algorithm on  an IEEE 123-node test feeder. The original test system is unbalanced. For all unbalanced case studies, we construct a balanced test system by ignoring the phase-to-phase admittance and replace all multi-phase lines with a one-phase line with average inductance and resistance of the phases. The  data for the test system can be found in \cite{feeder}.    We consider the  configuration, where all switched are open. The slack bus is the substation bus. The trading horizon is  one day with $H = 24$
time slots.       We add $10$ generators at different buses and  assume that each load aggregator serves between $100$ to $500$.  In Appendix F, we provide the simulation setup\cite{ont,toronto}.  
For the benchmark scenario, we consider
a system without demand response program for all load aggregators; thus, users operate their appliances right after they become awake. We perform
simulations using Matlab R2016b in a PC
with processor Intel(R) Core(TM) i7-3770K CPU@3.5 GHz.  




\subsubsection{Load aggregators' strategy} 
\begin{figure}[t]
\centering
\includegraphics[width=3.55in]{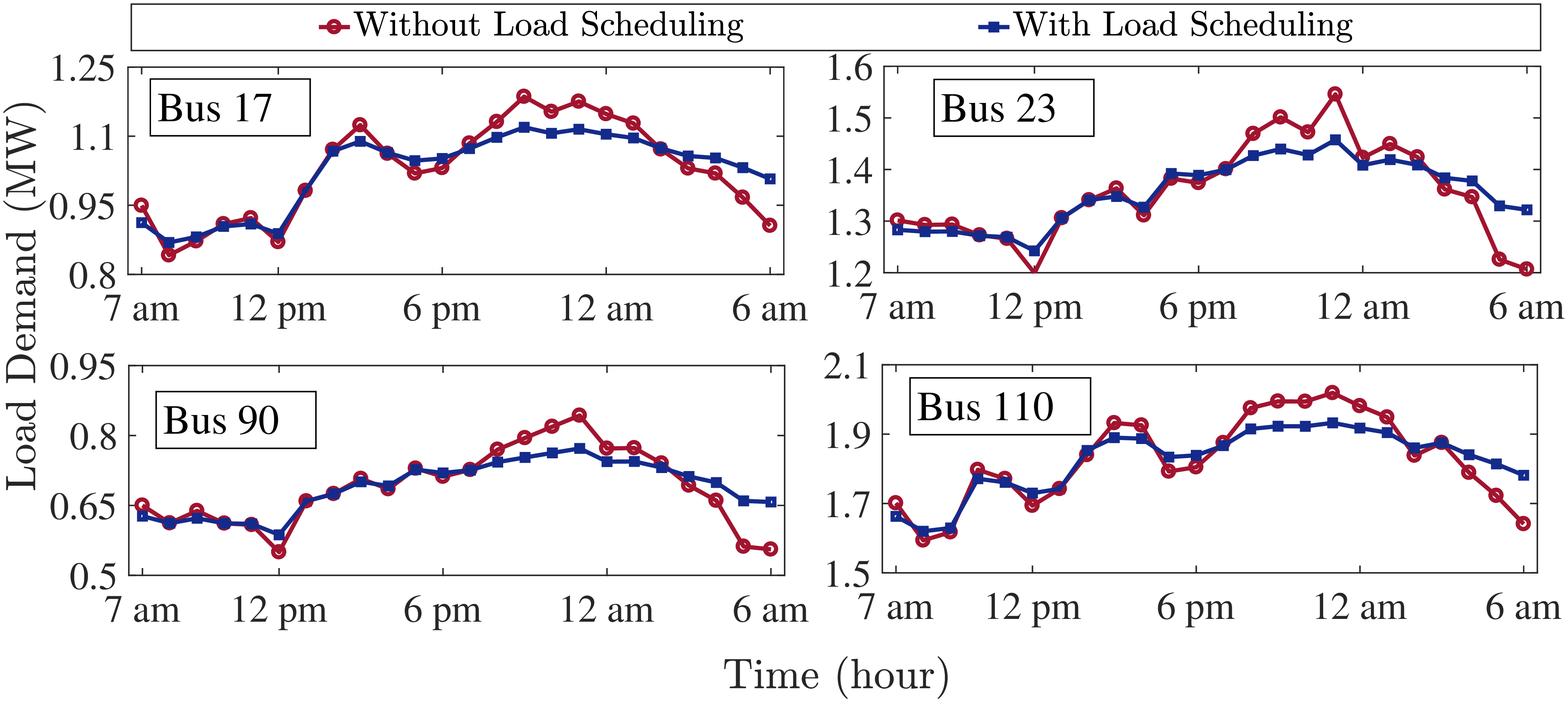}
\vspace{-7mm}
\caption{Load demand profiles over $24$ hours in buses $17$, $23$, $90$, and $110$  with and without appliances scheduling.}
\vspace{2mm}
\label{totload}
\centering
\includegraphics[width=3.55in]{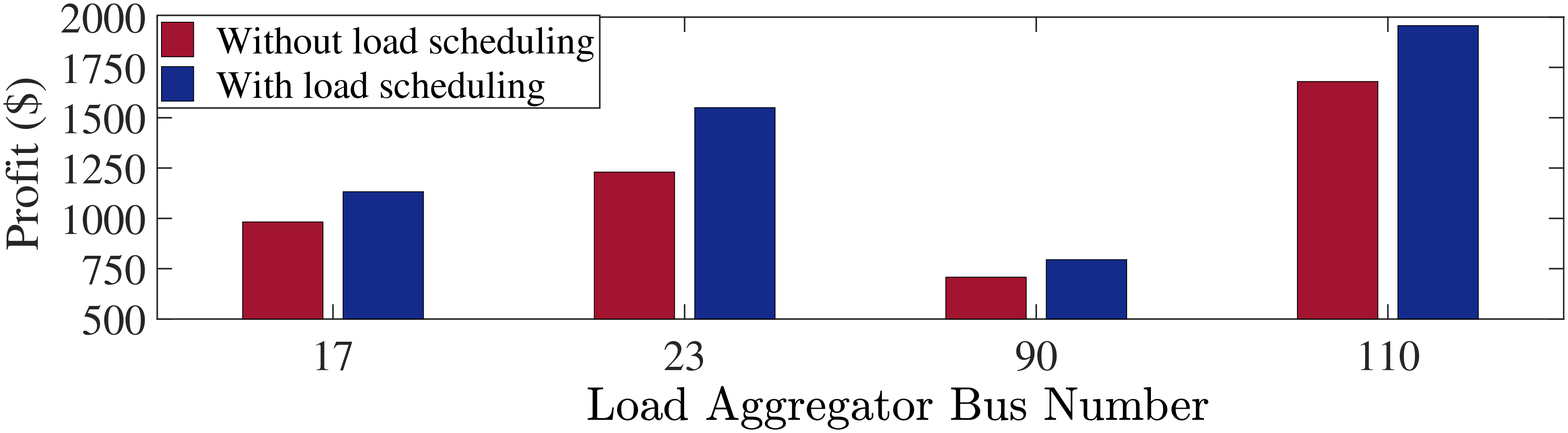}
\vspace{-20mm}
\caption{The profit for load aggregators with and without load scheduling.}
\label{aggcost}
\centering
\vspace{2mm}
\includegraphics[width=3.7in]{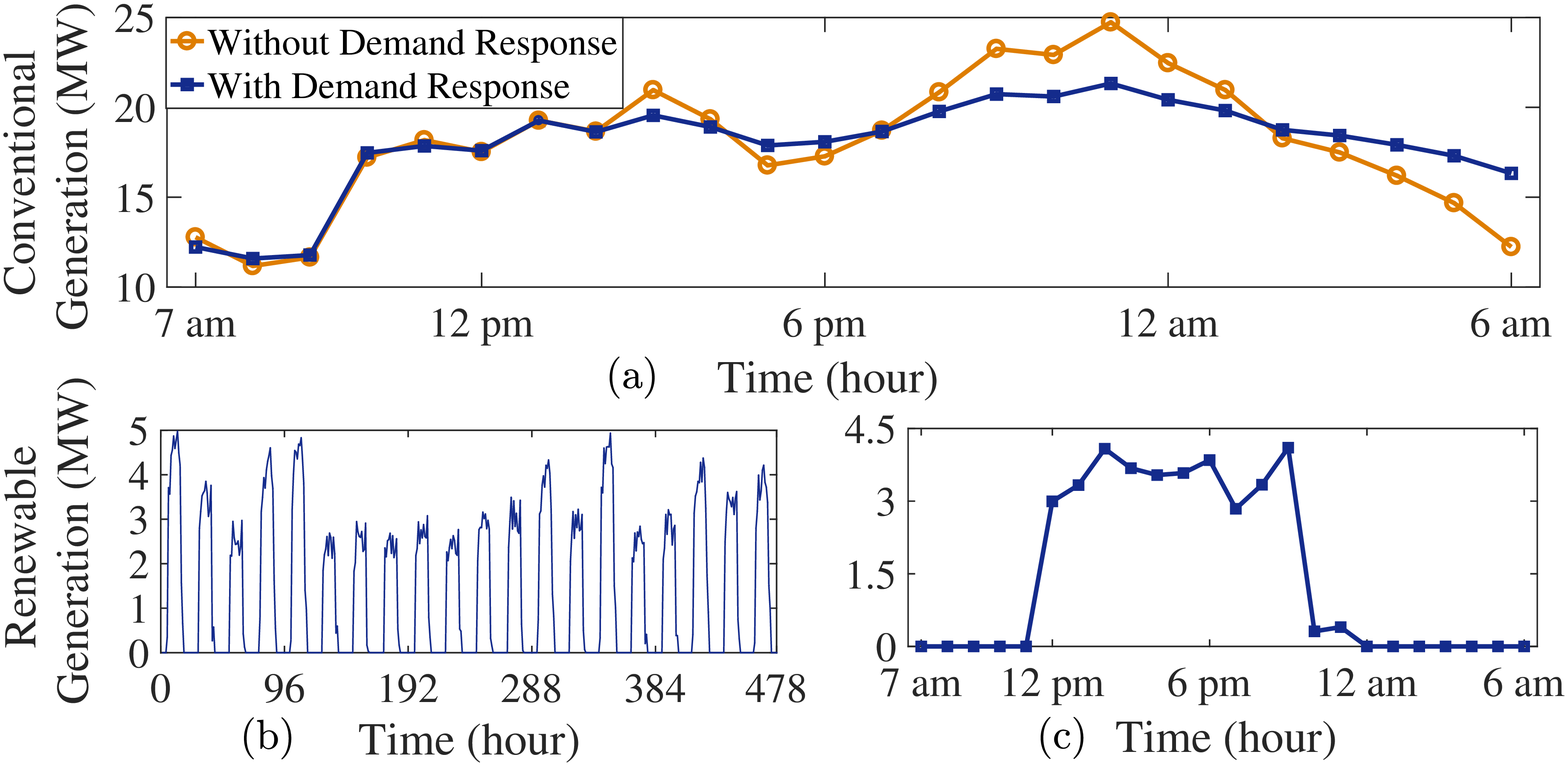}
\vspace{-4mm}
\caption{(a) The generation of the conventional unit of generator $15$. (b) The PV panel  historical data samples.  (c)  The offered output power of the PV panel   of generator $15$ in the market over the day. }
\label{gen1}
\vspace{-0mm}
\vspace{2mm}
\includegraphics[width=3.7in]{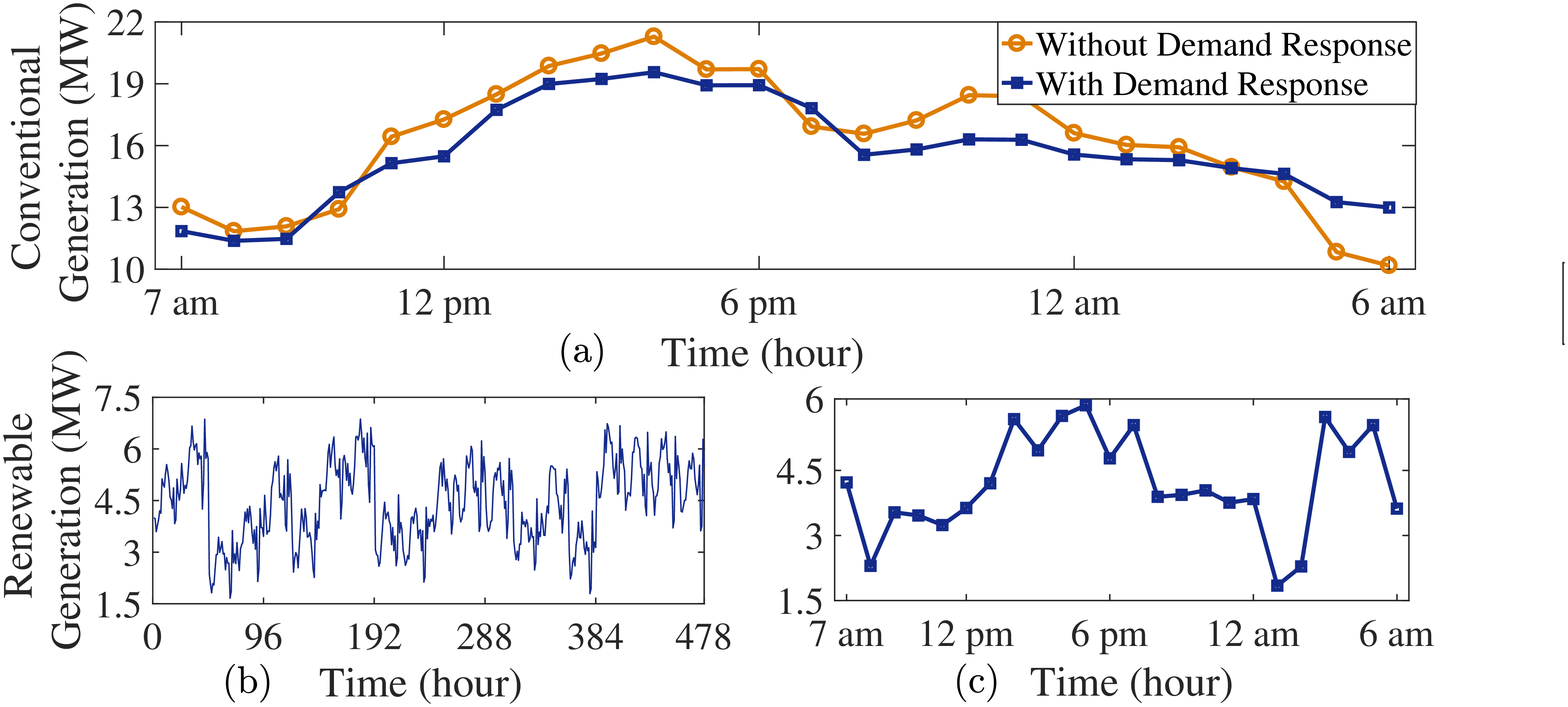}
\vspace{-4mm}
\caption{(a) The generation of the conventional unit of generator $60$. (b) The wind turbine  historical data samples.  (c) The presumed output power of the wind turbine   of generator $60$.}
\label{gen2}
\vspace{-0mm}
\end{figure}

Each  load aggregator executes Algorithm 1  to schedule the appliances of its users. Fig. \ref{totload} shows the load profile of the load aggregators in buses  $17$, $23$, $90$, and $110$  in the benchmark  scenario and the scenario  with load scheduling. Peak shaving can be observed in the load profiles. Since  Algorithm 1 is executed in real-time, the load aggregators can only modify the demand for upcoming time slots using the revealed information about the awake appliances in the current time slot and the estimated load demand for future time slots.  Results for \textit{all} load aggregators verify that by executing  Algorithm 1, the peak load demand is reduced by $14.5\%$ in average.  Load scheduling is performed by each load aggregator with the goal of increasing the profit in (\ref{costt}). Fig. \ref{aggcost} shows that the profit of the load aggregators $17$, $23$, $90$, and $110$  is increased. Specifically, results show that the profit for all load aggregators is increased by $17.8\%$ on average, since they can benefit from the price fluctuations by modifying the operation of their users' appliances.
\begin{figure}[t]
\centering
\includegraphics[trim={0cm 6.5cm 0cm 0},clip,width=3.7in]{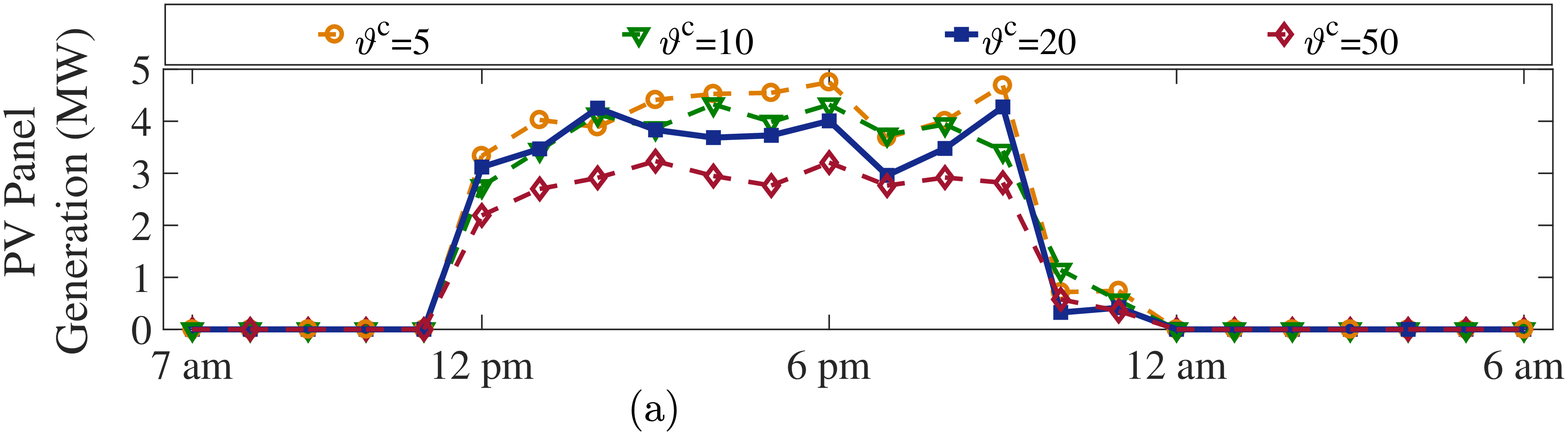}
\vspace{-17mm}
\includegraphics[width=3.7in]{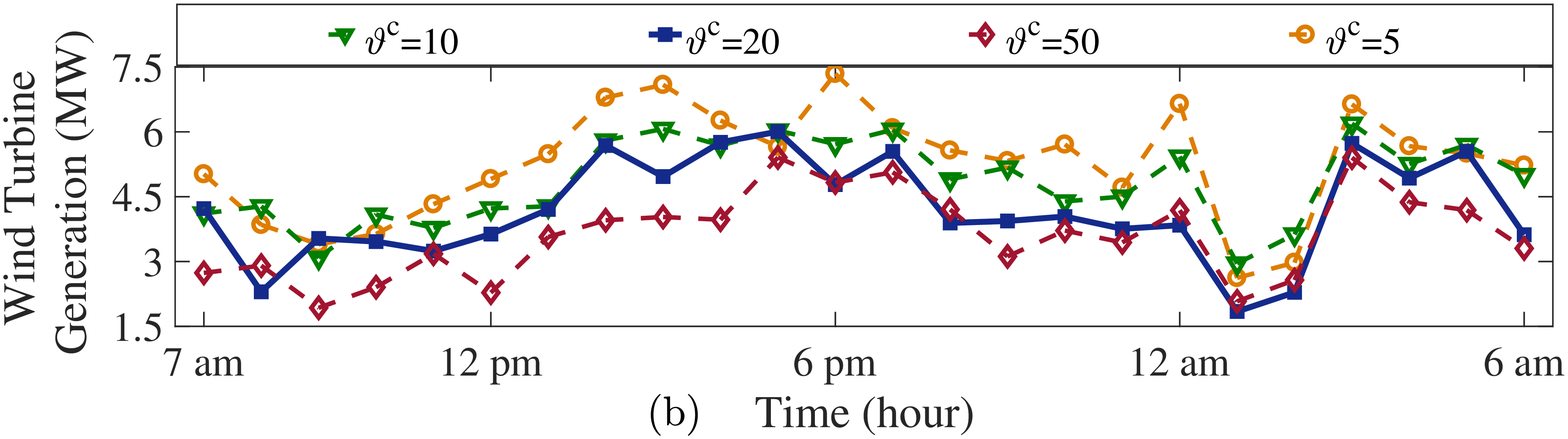}
\vspace{-5mm}
\caption{(a) The offered output power of the PV panel   of generator $15$ with different values of coefficient $\vartheta^{\text{c}}$. (b) The offered output power of the wind turbine   of generator $60$ with different values of coefficient $\vartheta^{\text{c}}$.}
\label{renewable}
\vspace{2mm}
\centering
\includegraphics[width=3.7in]{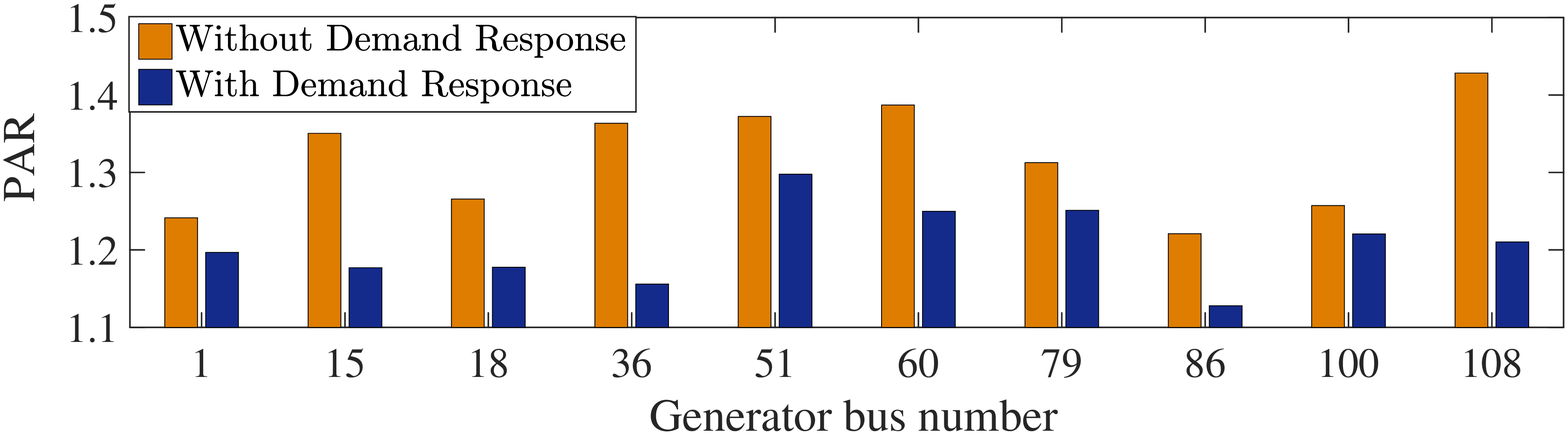}
\vspace{-23mm}
\caption{The PAR in the generation of the generators with and without demand response.}
\label{par}
\vspace{-2mm}
\end{figure}

\subsubsection{Generators' strategy}  On the other hand, generators can benefit from the users' load scheduling to reduce their peak generation, and thus their generation cost during peak hours. For example, Fig. \ref{gen1} (a) shows the active output power profile from the conventional unit of the generator in bus $15$. The peak generation level is reduced from $25$ MW to $20$ MW (i.e. $20\%$ reduction). Generator $15$ also has a PV panel with the historical generation record shown in Fig. \ref{gen1} (b). The generator executes Algorithm 1 and responds to the penalties $\bm{\beta}_j(t)$ from the DNO in each time slot to set the least-risk generation level for its PV unit. Fig. \ref{gen1} (c) shows the offers for generation $15$ over the day. Note that the offers may not be equal to the actual PV panel's generation in real-time, but the generation profile in Fig. \ref{gen1} (c) results in the optimal risk of energy shortage for PV panel in bus $15$. We also show the generation profile of the conventional unit of generator $60$ in Fig. \ref{gen2} (a). The reduction in peak generation can be observed. This generator has wind turbine with  the historical generation record shown in Fig. \ref{gen2} (b).  Fig. \ref{gen1} (c) shows the offers for wind turbine in bus $60$. 

The offers for  renewable units' generation mainly depends on the conservativeness of the DNO. In particular, when the weight coefficient $\vartheta^\text{c}$ in (\ref{ISO_obj}) is large, the DNO is risk-averse and forces the generators to prevent generation shortage in their renewable units. On the other hand, small coefficient $\vartheta^\text{c}$ means the DNO encourages the generators to offer higher amount of renewable generations. As an example, Figs. \ref{renewable} (a) and (b) show the renewable generation profiles of generators $15$ and $60$ for different values of $\vartheta^{\text{c}}$. When $\vartheta^{\text{c}}$ increases from $5$ to $50$, the generation levels decrease, since the penalties $\bm{\beta}_j(t)$ in (\ref{penal4}) increase. Hence, the generators offer lower renewable generations to reduce their cost of generation shortage.

\begin{figure}[t]
\centering
\includegraphics[width=3.7in]{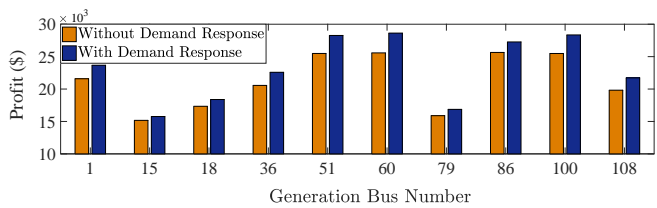}
\vspace{-20mm}
\caption{The profit of the generators with and without demand response.}
\label{profitgen}
\vspace{1mm}
\end{figure}

 To quantify the peak shaving, we consider the PAR of the generation. Fig. \ref{par} shows that the PAR is reduced for the generators by  $13\%$ on average. A lower PAR means a lower generation cost, and thus a higher profit. Fig.~\ref{profitgen} confirms that the generators' profit is increased by $10.3\%$ on average.

\subsubsection{Algorithm convergence}  
\begin{figure}[t]
\centering
\includegraphics[trim={0cm 6.1cm 0cm 0},clip,width=3.7in]{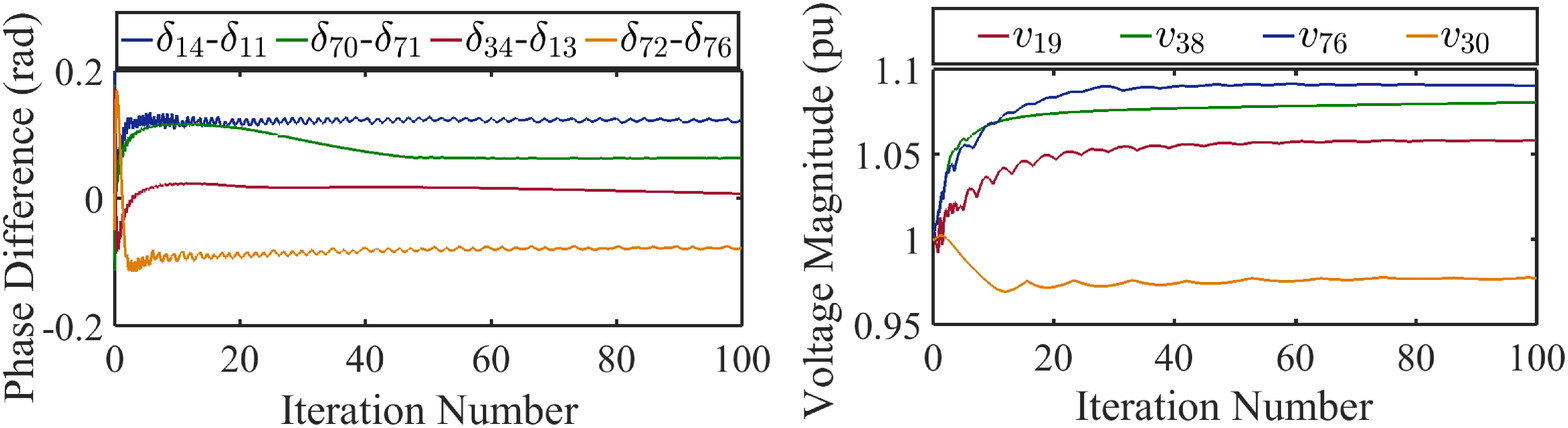}
\vspace{-8mm}
\caption{The convergence of phase differences and voltage magnitudes.}
\label{theta}
\vspace{2mm}
\centering
\includegraphics[width=3.7in]{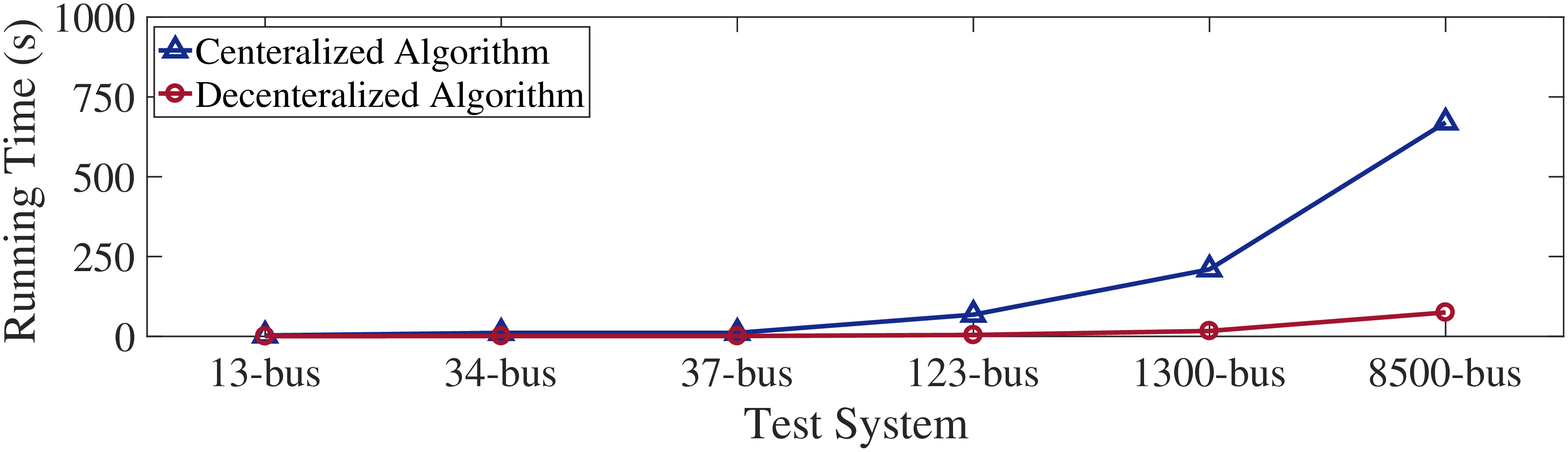}
\vspace{-22mm}
\caption{The running time of the centralized and decentralized algorithms.}
\label{run}
\vspace{3mm}
\centering
\includegraphics[width=3.7in]{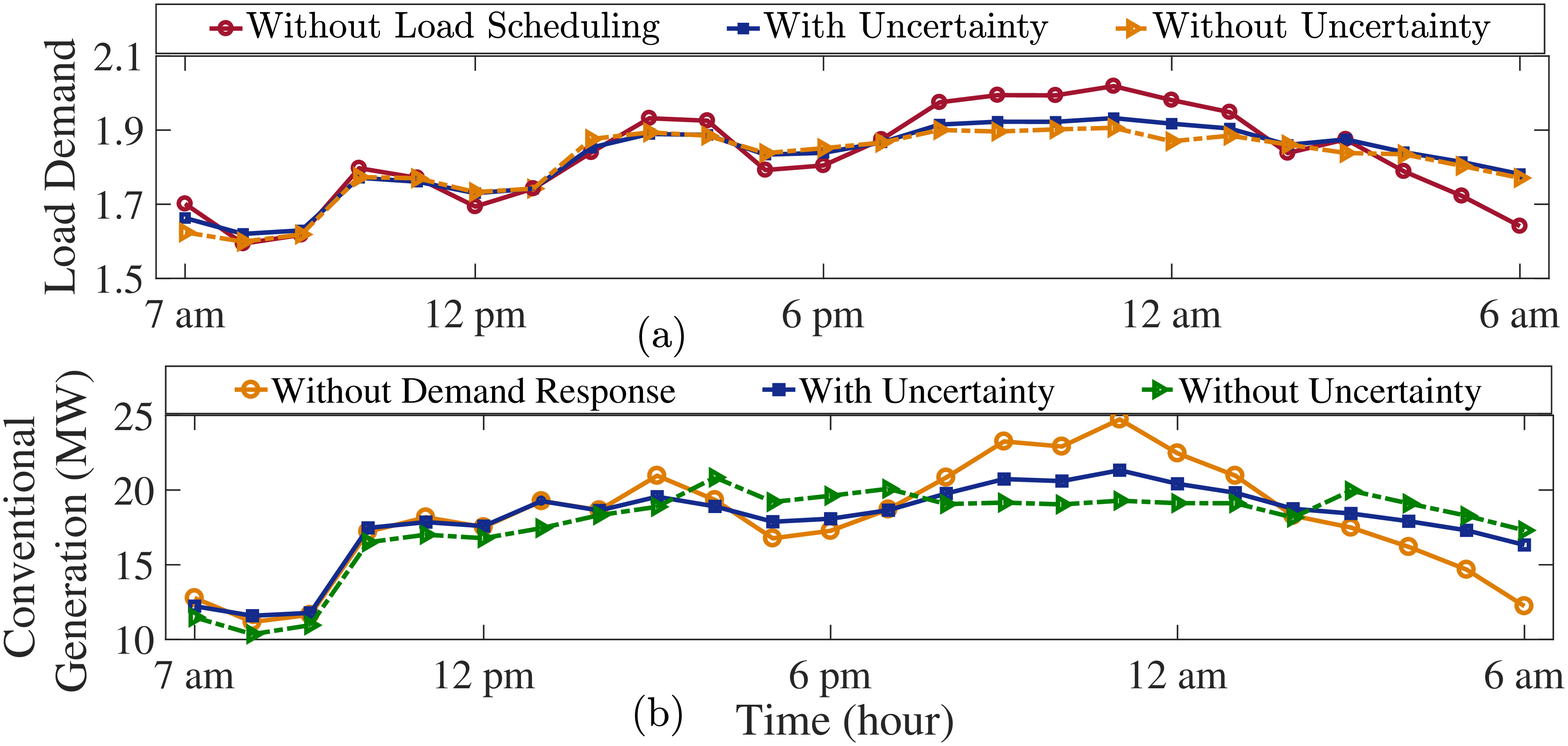}
\vspace{-7mm}
\caption{(a) Load scheduling with uncertainty and complete information. (b) Generation profile with uncertainty and complete information. }
\label{compare}
\end{figure}

We study the required number of iterations for convergence, which can be interpreted as an indicator of the number of  message exchanges among the load aggregators, generators and  DNO.   The  angle and magnitude of the voltage of the buses depend on all generators' and load aggregators' decision variables. Thus, the convergence of the these variables is a viable indicator of the convergence of all decision variables in  the system.  Since the values of the voltage angles of all buses can be added by a constant, we illustrate the convergence of the phase angle difference between the voltages of the buses at the end nodes of  the lines. As an example, we consider time slot $1$ and we provide the convergence of $\delta_{14}(1)-\delta_{11}(1)$, $\delta_{70}(1)-\delta_{71}(1)$, $\delta_{34}(1)-\delta_{13}(1)$,  and $\delta_{72}(1)-\delta_{76}(1)$ in Fig. \ref{theta} ($a$). We also show the voltage magnitude of the buses $19$, $38$, $76$, and $30$ in Fig. \ref{theta} ($b$). We can observe that $50$ iterations are enough for  convergence. The average running time of the algorithm for different initial conditions is $5$ seconds for $100$ random initial conditions. The low convergence rate and running time make the proposed algorithm  implementable for real-time interactions among entities in an energy market. 

 We also evaluate the running time of Algorithm 1 for larger test systems to show its scalability. Meanwhile, we compare its running time with a centralized algorithm, where the DNO solves problem problem~(\ref{ISO_problem}). We use MOSEK solver to solve the DNO's centralized problem~(\ref{ISO_problem}). The control signals in (\ref{penal1})$-$(\ref{penal4}) are determined in order to obtain the same solution for both the centralized and decentralized algorithms, and simulation results confirm this. We provide  the average running time of  Algorithm 1 and the centralized approach for six test systems (all  can be found in\cite{feeder} except the system with 1500 buses, which is a part of 8500-bus test system) in  Fig. \ref{run}. We can observe the the centralized algorithm suffer from a high running time due to a large number of decision variables and constraints. On the other hand,  Algorithm 1 is executed by each entity to solve its own optimization problem with its locally available information in a distributed fashion. Hence, the number of decision variables for each entity becomes independent of the size of the test system. The overall running time  of Algorithm 1 increases almost linearly with the number of buses due to the increase in the required number of iterations for convergence in larger test systems.

We also compare the performance of Algorithm 1 for the scenario with uncertainty in the load demand and renewable generations and  the scenario with complete information.  As an  example, we consider the load profile of the load aggregator $110$ and the generation profile of generator $15$ with a PV panel  in Figs. \ref{compare} $(a)$ and $(b)$.  The lack of  information makes the load aggregators more conservative, since it considers the \textit{worst-case} for the electric appliances in the upcoming time slots. Whereas, when the load aggregator has complete information, it can better manage the electric appliances especially during the peak hours. A lower peak  demand for the load aggregators results in a lower peak in the generation level of the conventional units. The conventional unit may  provide more power during the times when the PV panel has  high generation  (e.g. around 6 pm), as the PV generation  is known and the generator does not  take risk to offer a high renewable generation.

  \vspace{-2.2mm}
\section{Conclusion} \label{s6}

 In this paper, we proposed a real-time decentralized algorithm for energy trading among load aggregators and generators. {\color{black} Our proposed approach considers the uncertainty at both generation and demand sides.} In our model, the DNO sends control signals to the entities to encourage them towards  optimizing their objectives independently, while meeting the physical constraints of the power network. This study uses linearized ac optimal power flow formulation to increase the accuracy of the obtained solution. Further, we solve the problem in a real-time fashion to obtain the most recent optimal solution for each time-step.   To evaluate the performance of the proposed decentralized algorithm, we used an IEEE 123-bus test feeder connected to some renewable generators.  
 Although we considered the stochastic nature of renewables as well as ac power flow formulation, our algorithm converges in 50 iterations. We  evaluated the  price responsive load  profiles and generation values and showed that our method can benefit the load aggregators by increasing their profit by $17.8\%$, and the generators by reducing the PAR by $13\%$ and increasing their profit by $10.3\%$. Our algorithm benefit the DNO by maintaining the privacy issues and a lower computational time compared to the centralized approach.

\vspace{-3mm}
\bibliography{mybibfile4,mybibfile,ref,mybibfile2}

\begin{thebibliography}{10}
\providecommand{\url}[1]{#1}
\csname url@samestyle\endcsname
\providecommand{\newblock}{\relax}
\providecommand{\bibinfo}[2]{#2}
\providecommand{\BIBentrySTDinterwordspacing}{\spaceskip=0pt\relax}
\providecommand{\BIBentryALTinterwordstretchfactor}{4}
\providecommand{\BIBentryALTinterwordspacing}{\spaceskip=\fontdimen2\font plus
\BIBentryALTinterwordstretchfactor\fontdimen3\font minus
  \fontdimen4\font\relax}
\providecommand{\BIBforeignlanguage}[2]{{%
\expandafter\ifx\csname l@#1\endcsname\relax
\typeout{** WARNING: IEEEtran.bst: No hyphenation pattern has been}%
\typeout{** loaded for the language `#1'. Using the pattern for}%
\typeout{** the default language instead.}%
\else
\language=\csname l@#1\endcsname
\fi
#2}}
\providecommand{\BIBdecl}{\relax}
\BIBdecl

\bibitem{Bo}
B.~Chai, J.~Chen, Z.~Yang, and Y.~Zhang, ``Demand response management with
  multiple utility companies: {A} two-level game approach,'' \emph{IEEE Trans.
  on Smart Grid}, vol.~5, no.~2, pp. 722--731, Mar. 2014.

\bibitem{maharjan}
S.~Maharjan, Q.~Zhu, Y.~Zhang, S.~Gjessing, and T.~Basar, ``Dependable demand
  response management in the smart grid: {A} stackelberg game approach,''
  \emph{IEEE Trans. on Smart Grid}, vol.~4, no.~1, pp. 120--132, Mar. 2013.

\bibitem{deng}
R.~Deng, Z.~Yang, F.~Hou, M.-Y. Chow, and J.~Chen, ``Distributed real-time
  demand response in multiseller-multibuyer smart distribution grid,''
  \emph{IEEE Trans. on Power Systems}, vol.~PP, no.~99, pp. 1--11, Oct. 2014.

\bibitem{farhad}
F.~Kamyab \emph{et~al.}, ``Demand response program in smart grid using supply
  function bidding mechanism,'' \emph{IEEE Trans. on Smart Grid}, vol.~7,
  no.~3, pp. 1277 -- 1284, 2016.

\bibitem{ucdr4}
M.~Parvania, M.~Fotuhi-Firuzabad, and M.~Shahidehpour, ``{ISO's} optimal
  strategies for scheduling the hourly demand response in day-ahead markets,''
  \emph{IEEE Trans. on Power Systems}, vol.~29, no.~6, pp. 2636--2645, 2014.

\bibitem{DDCOPF1}
L.~Gan, N.~Li, U.~Topcu, and S.~H. Low, ``Exact convex relaxation of optimal
  power flow in radial networks,'' \emph{IEEE Trans. on Automatic Control},
  vol.~60, no.~1, pp. 72--87, 2015.

\bibitem{opfdr1}
W.~Shi, N.~Li, X.~Xie, C.~Chu, and R.~Gadh, ``Optimal residential demand
  response in distribution networks,'' \emph{IEEE Journal on Selected Areas in
  Comm.}, vol.~32, no.~7, pp. 1441--1450, Jun. 2014.

\bibitem{opfdr3}
N.~Li, L.~Gan, L.~Chen, and S.~Low, ``An optimization-based demand response in
  radial distribution networks,'' in \emph{Proc. of IEEE Globecom}, Anaheim,
  CA, Anaheim, CA 2012.

\bibitem{new1}
E.~Dall'Anese, H.~Zhu, and G.~B. Giannakis, ``Distributed optimal power flow
  for smart microgrids,'' \emph{IEEE Trans. on Smart Grid}, vol.~4, no.~3, pp.
  1464--1475, Sept. 2013.

\bibitem{new2}
A.~G. Bakirtzis and P.~N. Biskas, ``A decentralized solution to the {DC-OPF} of
  interconnected power systems,'' \emph{IEEE Trans. on Power Systems}, vol.~18,
  no.~3, pp. 1007--1013, Aug. 2003.

\bibitem{new4}
T.~Erseghe, ``Distributed optimal power flow using {ADMM},'' \emph{IEEE Trans.
  on Power Systems}, vol.~29, no.~5, pp. 2370--2380, Sept. 2014.

\bibitem{new6}
Q.~Peng and S.~H. Low, ``Distributed algorithm for optimal power flow on a
  radial network,'' in \emph{Proc. of IEEE Conf. on Decision and Control}, Dec.
  2014, pp. 167--172.

\bibitem{new8}
S.~Magnusson, P.~C. Weeraddana, and C.~Fischione, ``A distributed approach for
  the optimal power-flow problem based on {ADMM} and sequential convex
  approximations,'' \emph{IEEE Trans. on Control of Network Systems}, vol.~2,
  no.~3, pp. 238--253, Sept. 2015.

\bibitem{new5}
J.~M. Arroyo and F.~D. Galiana, ``Energy and reserve pricing in security and
  network-constrained electricity markets,'' \emph{IEEE Trans. on Power
  Systems}, vol.~20, no.~2, pp. 634--643, May 2005.

\bibitem{newdr}
S.~Mhanna, A.~C. Chapman, and G.~Verbi{\v{c}}, ``A fast distributed algorithm
  for large-scale demand response aggregation,'' \emph{IEEE Trans. on Smart
  Grid}, vol.~7, no.~4, pp. 2094--2107, 2016.

\bibitem{online1}
M.~J. Dolan, E.~M. Davidson, I.~Kockar, G.~W. Ault, and S.~D. McArthur,
  ``Distribution power flow management utilizing an online optimal power flow
  technique,'' \emph{IEEE Trans. on Power Systems}, vol.~27, no.~2, pp.
  790--799, May 2012.

\bibitem{online3}
E.~Belic, N.~Lukac, K.~Dezelak, B.~Zalik, and G.~Stumberger, ``Gpu-based online
  optimization of low voltage distribution network operation,''
  \emph{\rm{accepted for publication in} \it{IEEE Trans. on Smart Grid}}, 2017.

\bibitem{online4}
L.~Gan and S.~H. Low, ``An online gradient algorithm for optimal power flow on
  radial networks,'' \emph{IEEE J. on Selected Areas in Comm.}, vol.~34, no.~3,
  pp. 625--638, Mar. 2016.

\bibitem{online5}
A.~Hauswirth, S.~Bolognani, G.~Hug, and F.~D{\"o}rfler, ``Projected gradient
  descent on {Riemannian} manifolds with applications to online power system
  optimization,'' in \emph{Proc. of Allerton Conf. on Communications, Control
  and Computing}, Sept. 2016, pp. 225--232.

\bibitem{online6}
S.-J. Kim, G.~B. Giannakis, and K.~Y. Lee, ``Online optimal power flow with
  renewables,'' in \emph{Proc. of Asilomar Conf. on Signals, Systems and
  Computers}, Nov. 2014, pp. 355--360.

\bibitem{online7}
D.~Mehta, A.~Ravindran, B.~Joshi, and S.~Kamalasadan, ``Graph theory based
  online optimal power flow control of power grid with distributed flexible ac
  transmission systems (d-facts) devices,'' in \emph{Proc. of North American
  Power Symposium (NAPS)}, Oct. 2015, pp. 1--6.

\bibitem{low}
N.~Li, L.~Chen, and S.~H. Low, ``Optimal demand response based on utility
  maximization in power networks,'' in \emph{IEEE Power and Energy Society
  General Meeting}, Jul. 2011, pp. 1--8.

\bibitem{genfunc}
T.~Li and M.~Shahidehpour, ``Price-based unit commitment: a case of lagrangian
  relaxation versus mixed integer programming,'' \emph{IEEE Trans. on Power
  Systems}, vol.~20, no.~4, pp. 2015--2025, 2005.

\bibitem{adaptive}
D.~Bertsimas, E.~Litvinov, X.~A. Sun, J.~Zhao, and T.~Zheng, ``Adaptive robust
  optimization for the security constrained unit commitment problem,''
  \emph{IEEE Trans.s on Power Systems}, vol.~28, pp. 52--63, Feb. 2013.

\bibitem{new7}
G.~Hug-Glanzmann and G.~Andersson, ``Decentralized optimal power flow control
  for overlapping areas in power systems,'' \emph{IEEE Trans. on Power
  Systems}, vol.~24, no.~1, pp. 327--336, Feb. 2009.

\bibitem{boyd2004convex}
S.~Boyd and L.~Vandenberghe, \emph{{Convex Optimization}}.\hskip 1em plus 0.5em
  minus 0.4em\relax Cambridge University Press, 2004.

\bibitem{feeder}
\BIBentryALTinterwordspacing
 [Online]. Available: \url{https://ewh.ieee.org/soc/pes/dsacom/testfeeders/}
\BIBentrySTDinterwordspacing

\bibitem{ont}
\BIBentryALTinterwordspacing
{Independent Electricty System Operator (IESO)}. [Online]. Available:
  \url{http://www.ieso.ca}
\BIBentrySTDinterwordspacing

\bibitem{toronto}
\BIBentryALTinterwordspacing
 [Online]. Available:
  \url{http://www.torontohydro.com/sites/electricsystem/\\residential/yourbilloverview/Pages/ApplianceChart.aspx}
\BIBentrySTDinterwordspacing

\end{thebibliography}

 \bibliographystyle{IEEEtran}
\end{document}